\documentclass[aps,pra,showpacs,showkeys,superscriptaddress,twocolumn]{revtex4}
\usepackage{amsmath}
\usepackage{graphicx}

\begin{document}

\title{Quantum Decoherence of Charge Qubit coupled to Nonlinear Nanomechanical
Resonator}

\author{CHEN Cheng}
\affiliation{College of Applied Sciences, Beijing University of Technology, Beijing,
100124, China}

\author{GAO Yi-bo}
\email{ybgao@bjut.edu.cn}
\affiliation{College of Applied Sciences, Beijing University of Technology, Beijing,
100124, China}

\begin{abstract}
When the nonlinearity of nanomechanical resonator is not negligible,
the quantum decoherence of charge qubit is studied analytically. Using
nonlinear Jaynes-Cummings model, one explores the possibility of being
quantum data bus for nonlinear nanomechanical resonator, the nonlinearity
destroys the dynamical quantum information-storage and maintains the
revival of quantum coherence of charge qubit. With the calculation
of decoherence factor, we demonstrate the influence of the nonlinearity
of nanomechanical resonator on engineered decoherence of charge qubit.
\end{abstract}

\maketitle

\section{introduction}

Any quantum system, is immersed into the environment, can not be isolated
from the environment completely.~\cite{Zurek83}In literatures, one
mimics the environment with the ``bath'' model. The fluctuation
of the bath induces the dissipation and decoherence of the quantum
system such that long decoherence time is very important for qubit.
Actually, to implement quantum computation, one often uses a longer
life-time medium to store the state of qubit.~\cite{DiVincenzo00} Now
quantum decoherence and quantum information-storage have been the
central issues in the study of quantum information and computation.

Both long decoherence-time qubit and long life-time medium are potential
candidates for quantum information. The traditional cavity QED system,~\cite{Raimond01}
consisting of two-level atom and microwave cavity, can implement this
quantum information-storage process. Now the decoherence time of Josephson
junction qubit has been of order 10$\mu s$.~\cite{Schoelkopf11}
Josephson junction qubit~\cite{Makhlin01} can be coupled to superconducting
microwave cavity~\cite{Yang04}and transmission line resonator~\cite{Blais04}.
Recently, the integration of Josephson junction qubit and nanomechanical
resonators~\cite{Xue07} are attracting considerable attentions. These
nanomechanical resonators can be easily fabricated using technologies
of nanoelectronics (see, e.g., Refs~\cite{Roukes09,Roukes13}). The
dynamics of all these coupled systems could be described by the Jaynes-Cummings
Hamiltonian.~\cite{Jaynes63}

In contrast to nanomechanical resonator being assumed as a harmonic oscillator, the nonlinearity of nanomechanical resonator has attracted
more and more attentions,~\cite{Roukes13} such as one can enhance
the anharmonicity of nanomechanical resonators by subjecting them
to inhomogenous electrostatic fields,~\cite{Hartmann12} generating
Yurke-Stoler states.~\cite{Milburn09} Here the coupled system including
charge qubit and nonlinear nanomechanical resonator is considered.

This paper is organized as follows. In Sec.~II, we describe the proposed
model of a charge qubit interacting with a nonlinear nanomechanical
resonator. In Sec.~III, the nonlinearity of nanomechanical is considered
in implementing the process of dynamical quantum information-storage.
In Sec.~VI, the influence of the nonlinearity of nanomechanical resonator
on engineered decoherence is studied analytically.

\begin{figure}[th]
\includegraphics[bb=101 71 522 327,width=8cm,clip]{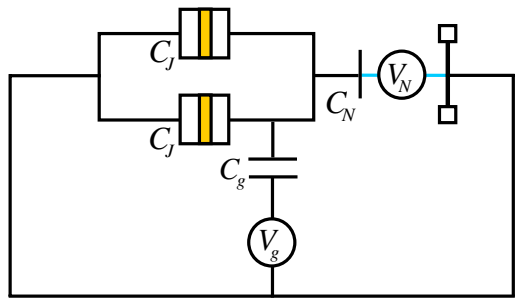}
\caption{Schematic diagram of a charge qubit capacitively coupled to a nanomechanical resonator. An externally applied voltage on the capacitor ($C_{N}$) formed by the superconducting island of Josephson junction and the nanomechanical resonator. Here, $C_{J}$ and $C_{g}$ represent the capacitances for the Josephson junctions and the gate capacitor respectively. $V_{g}$ is the gate
voltage applied to charge qubit via the gate capacitor, and $V_{N}$ is an externally applied voltage on the nanomechanical resonator.}
\label{setup}
\end{figure}

\section{model}

Due to the nonlinearity, the nanomechanical resonator is not assumed
as harmonic resonator. The Hamiltonian for nonlinear nanomechanical
resonator is described by~\cite{Duffing01}
\begin{equation}
H_{\mathrm{N}}=\frac{p^{2}}{2m}+\frac{1}{2}m\omega^{2}x^{2}+\frac{1}{4}\alpha x^{4},\label{Ham-namr}
\end{equation}
where $m$ is the effective mass, $\omega$ is the linear resonator
frequency and the parameter $\alpha$ measures the strength of nonlinearity.
The canonical coordinate $x$ and momentum $p$ satisfies the commutation
relation $\left[x,p\right]=i$. Hereafter, we assume $\hbar=1$. The
$x$ and $p$ can be also represented by the annihilation and creation
operators ($a$ and $a^{\dagger}$),
\begin{eqnarray}
x & = & \frac{1}{\sqrt{2m\omega}}\left(a^{\dag}+a\right),\label{eq:2-1}\\
p & = & i\sqrt{\frac{m\omega}{2}}\left(a^{\dag}-a\right).\label{eq:2-2}
\end{eqnarray}
Thus the Hamiltonian in Eq.~(\ref{Ham-namr}) can be rewritten as
\begin{equation}
H_{N}=\omega a^{\dagger}a+\frac{\chi}{6}\left(a^{\dag}+a\right)^{4}\label{eq:Ham-N}
\end{equation}
with the nonlinear parameter is $\chi=\alpha/\left(8m\omega\right)$.

As shown in Fig.~1, with the oscillation of nanomechanical resonator,
the Cooper pair box (charge qubit) is capacitively coupled to the
nanomechanical resonator. Here $C_{N}(x)$ denotes the capacitance
between nanomechanical resonator and the superconducting island of
Cooper pair box. The distance $d$ between nanomechanical resonator
and superconducting island is assumed much larger than the amplitude
$x$ of the oscillation of nanomechanical resonator, i.e., $d\gg x$.
In this case, one can approximately simplify the capacitance $C_{N}(x)$
as
\begin{equation}
C_{N}(x)\simeq C_{N}\left(1-\frac{x}{d}\right).
\end{equation}

According to the results in Refs.~\cite{EPJB04,Gao09}, the Hamiltonian
of charge qubit reads
\begin{equation}
H_{c}=\frac{\omega_{q}}{2}\sigma_{z}+g\left(a+a^{\dagger}\right)\sigma_{x}\label{Ham-c2}
\end{equation}
with respect to the qubit frequency $\omega_{a}=E_{J}$ and the coupling
constant $g=4E_{C}n_{N}x_{zp}/d$. These spin operators $\{\sigma_{z},$
$\sigma_{\pm}\}$ are defined in the basis of charge qubit $\left\{ \vert0\rangle_{q},\vert1\rangle_{q}\right\} $,
i.e.,
\begin{equation}
\sigma_{+}\left\vert 0\right\rangle _{q}=\left\vert 1\right\rangle _{q},\ \ \sigma_{-}\left\vert 1\right\rangle _{q}=\left\vert 0\right\rangle _{q}.
\end{equation}
Notice that the coupling constant $g$ is determined by the bias voltage $V_{N}$ ($\propto n_{N}$), and one can tune the qubit frequency $\omega_{q}$ through the external magnetic field. Then the total Hamiltonian
for charge qubit and nonlinear nanomechanical resonator is
\begin{equation}
H=H_{c}+H_{\mathrm{N}}.\label{Ham-total}
\end{equation}
Where $H_{c}$ and $H_{\mathrm{N}}$ are given by
Eqs.~(\ref{eq:Ham-N},\ref{Ham-c2}) respectively.

\section{revival of quantum coherence}
During the process of the preparation, manipulation and measurement
of quantum state, the loss and revival of quantum coherence is very important for building a real quantum computer. With a longer life-time medium (data
bus), one can store, manipulate and communicate quantum information.~\cite{DiVincenzo00}
Considering the nonlinearity of nanomechanical resonator, the possibility
of being quantum data bus in quantum computation will be explored
in this section.

One can control the Josephson coupling energy $E_{J}$ such that the
qubit is resonant with the nanomechanical resonator ($\omega_{q}=\omega$).
Using the rotating-wave approximation, the total Hamiltonian in Eq.~(\ref{Ham-total}) becomes
\begin{equation}
H=\frac{\omega}{2}\sigma_{z}+g\left(a^{\dagger}\sigma_{-}+a\sigma_{+}\right)+\left(\omega+\chi\right)a^{\dagger}a+\chi\left(a^{\dagger}a\right)^{2}.\label{Ham-RWA}
\end{equation}
It is just the same as the nonlinear Jaynes-Cummings Hamiltonian.~\cite{NJC92} The nonlinear part in the above Hamiltonian $\chi\left(a^{\dagger}a\right)^{2}$
is obtained from the quartic potential ($\sim x^{4}$ seen in Eq.~(\ref{eq:Ham-N})). Here the nonlinear parameter $\chi$ is assumed
as much smaller than the coupling constant $g$. When the nonlinear
parameter $\chi$ vanishes, the Hamiltonian in Eq.~(\ref{Ham-RWA})
reduces into the Jaynes-Cummings Hamiltonian.~\cite{Jaynes63} In the
following, the influence of the nonlinear part $\chi\left(a^{\dagger}a\right)^{2}$
on quantum information storage and quantum coherence will be analyzed
carefully.

Corresponding to the Hamiltonian in Eq.~(\ref{Ham-RWA}), we can solve
the eigenvalues
\begin{equation}
E_{\pm}=\left(\frac{\omega}{2}+\chi\right)\pm g_{\chi}\label{eigenvalue}
\end{equation}
 and the eigenstates
\begin{eqnarray*}
\vert+\rangle & = & \cos\frac{\theta}{2}\vert01\rangle+\sin\frac{\theta}{2}\vert10\rangle,\\
\vert-\rangle & = & -\sin\frac{\theta}{2}\vert01\rangle+\cos\frac{\theta}{2}\vert10\rangle.
\end{eqnarray*}
Where some parameters are defined as $g_{\chi}=\sqrt{g^{2}+\chi^{2}}$
and $\theta=\arcsin\left(g/g_{\chi}\right)$. Here the nonlinear parameter
$\chi$ in Eq.~(\ref{eigenvalue})) shows that the nonlinearity of
nanomechanical resonator modifies the Rabi oscillation frequency $g_{\chi}$
in Jaynes-Cummings model.

Now the initial state of charge qubit is prepared in the coherent
superposition state, $\alpha\left\vert 0\right\rangle _{q}+\beta\left\vert 1\right\rangle _{q}$,
and the nanomechanical resonator in the vacuum state $\vert0\rangle_{N}$.
Then the initial state of the total system writes
\[
\left\vert \psi\left(0\right)\right\rangle =\left(\alpha\left\vert 0\right\rangle _{q}+\beta\left\vert 1\right\rangle _{q}\right)\otimes\left\vert 0\right\rangle _{N}
\]
for $\vert\alpha\vert^{2}+\vert\beta\vert^{2}=1$ and $\vert0\rangle_{q}\otimes\vert0\rangle_{N}=\vert00\rangle$.
When the charge qubit begins to interact with the nanomechanical resonator,
the state of the total system evolves into
\begin{eqnarray*}
\left\vert \psi\left(t\right)\right\rangle  & = & C_{00}\left(t\right)\left\vert 00\right\rangle +C_{01}\left(t\right)\vert01\rangle+C_{10}\left(t\right)\vert10\rangle.
\end{eqnarray*}
Some parameters are given in the following
\begin{eqnarray}
C_{00}\left(t\right) & = & \alpha e^{-i\frac{\omega}{2}t},\nonumber \\
C_{01}\left(t\right) & = & -i\beta \sin\theta e^{-i\left(\frac{\omega}{2}+\chi\right)}\sin\left(g_{\chi}t\right),\nonumber \\
C_{10}\left(t\right) & = & \beta e^{-i\left(\frac{\omega}{2}+\chi\right)}\left(\cos\left(g_{\chi}t\right)+i\cos\theta \sin\left(g_{\chi}t\right)\right).\label{parameters}
\end{eqnarray}
There exists the condition $\vert C_{00}\vert^{2}+\vert C_{01}\vert^{2}+\vert C_{10}\vert^{2}=1.$

The process of quantum information-storage can be implemented by
\begin{equation}
\left(\alpha\left\vert 0\right\rangle _{q}+\beta\left\vert 1\right\rangle _{q}\right)\otimes\left\vert 0\right\rangle _{N}\rightarrow\left\vert 0\right\rangle _{q}\otimes\left(\alpha\left\vert 0\right\rangle _{N}+\beta\left\vert 1\right\rangle _{N}\right).\label{eq:information transfer}
\end{equation}
Considering the nonlinearity, the time evolution of the probability
of quantum information transferred from qubit to nanomechanical resonator
is
\begin{eqnarray}
P & = & \vert C_{00}\left(t\right)\vert^{2}+\left\vert C_{01}\left(t\right)\right\vert ^{2}\nonumber \\
 & = & \vert\alpha\vert^{2}+\vert\beta\vert^{2}\left[\sin\theta \sin\left(g_{\chi}t\right)\right]^{2}.\label{probability}
\end{eqnarray}
When the nonlinearity is not considered in Eqs.~(\ref{Ham-RWA},\ref{parameters})
($\chi=0$), the probability $P$ is 1. Then the quantum information
can be transferred from charge qubit to nanomechanical resonator.
Obviously seen in Eq.~(\ref{parameters}), the probability $P$ is
smaller than $1$. Then the process of quantum information-storage
can not be implemented and the nonlinear nanomechanical resonator
can not be a quantum data bus.

At time $\tau=k\pi/g_{\chi}$ for $k$s being integers, there exists
$C_{01}\left(\tau\right)=0$ and the state of the total system evolves
into
\begin{equation}
\left\vert \psi\left(\tau\right)\right\rangle =\left(C_{00}\left(\tau\right)\left\vert 0\right\rangle _{q}+C_{10}\left(\tau\right)\vert1\rangle_{q}\right)\otimes\vert0\rangle_{N}.\label{revival}
\end{equation}
Here the quantum information ($\alpha$ and $\beta$) returns into
the qubit and the nonlinearity maintains the revival of quantum coherence of charge qubit.

\section{engineered decoherence}

In previous work,~\cite{Raimond01} the traditional cavity QED system
(consisting of a two-level atom and a microwave cavity) demonstrates
a reversible decoherence process of mesoscopic superposition of field
states. This model (consisting of a charge qubit and a nonlinear nanomechanical
resonator) will show the influence of the nonlinearity on its engineered
decoherence.

According to Eq.~(\ref{Ham-total}), we turn off the Josephson coupling energy $E_{J}$, i.e., $\omega_{q}=0$. Thus the Hamiltonian corresponding to a standard quantum measurement model is obtained,~\cite{EPJB04}
\[
H=H_{0}\left\vert 0\right\rangle \left\langle 0\right\vert +H_{1}\left\vert 1\right\rangle \left\langle 1\right\vert .
\]
where the effective Hamiltonian $H_{k}$ is
\begin{equation}
H_{k}=\left(-1\right)^{k}g\left(a^{\dagger}+a\right)+\omega a^{\dagger}a+\frac{\chi}{6}\left(a+a^{\dagger}\right)^{4}\label{eq:driven kerr Ham}
\end{equation}
with respect to the qubit state $\left\vert k\right\rangle $ (for
$k=0,1$) which is the eigenvectors of spin operator $\sigma_{x}$.
Obviously the Hamiltonian $H_{k}$ in Eq.~(\ref{eq:driven kerr Ham})
describes a driven nonlinear oscillator.

Assuming the initial state of the nanomechanical resonator in vacuum
state and the qubit in a superposition state, then the initial state
of the total system is
\[
\left\vert \psi\left(0\right)\right\rangle =\left(\alpha\left\vert 0\right\rangle +\beta\left\vert 1\right\rangle \right)\otimes\left\vert 0\right\rangle_{N} .
\]
Here the time evolution of wave function for the total system is
\begin{eqnarray}
\left\vert \psi\left(t\right)\right\rangle  & = & e^{-iHt}\left\vert \psi\left(0\right)\right\rangle \nonumber \\
 & = & \alpha\left\vert 0\right\rangle \otimes\left\vert \mu_{0}\left(t\right)\right\rangle +\beta\left\vert 1\right\rangle \otimes\left\vert \mu_{1}\left(t\right)\right\rangle \label{eq:time evolution}
\end{eqnarray}
where we have defined
\begin{equation}
\left\vert \mu_{k}\left(t\right)\right\rangle =e^{-iH_{k}t}\left\vert 0\right\rangle_{N} .\label{eq:mu-k}
\end{equation}
For any coherent state, we have
\[
\left\vert \alpha\right\rangle =e^{\alpha a^{\dagger}-\alpha^{*}a}\left\vert 0\right\rangle_{N} .
\]
Applying the coherent displacement transformation
\[
D\left(\alpha_{k}\right)=e^{\alpha_{k}a^{\dagger}-\alpha_{k}^{*}a}
\]
 and neglecting the fast oscillating terms, the Hamiltonian $H_{k}$
in Eq.~(\ref{eq:driven kerr Ham}) is transformed into the effective
Hamiltonian
\[
H_{k}^{eff}=D^{\dagger}\left(\alpha_{k}\right)H_{k}D\left(\alpha_{k}\right)=\Omega a^{\dagger}a+\chi\left(a^{\dagger}a\right)^{2}
\]
where the transformed angular frequency of nanomechanical resonator
is $\Omega=\omega+\chi+8\lambda^{2}\chi$ for $\alpha_{k}=\left(-1\right)^{k}\lambda=\left(-1\right)^{k}g/\left(\omega+\chi\right)$.
The state vector in Eq.~(\ref{eq:mu-k}) writes
\begin{equation}
\left\vert \mu_{k}\left(t\right)\right\rangle =D\left(\alpha_{k}\right)e^{-iH_{k}^{eff}t}D^{\dagger}\left(\alpha_{k}\right)\left\vert 0\right\rangle_{N} .\label{mu-k}
\end{equation}

The process of quantum decoherence could be described by time evolution
of the reduced density matrix of quantum system and the off-diagonal
elements of reduced density matrix measures the decoherence. According
to Eq.~(\ref{eq:time evolution}), the time evolution of density matrix
for the total system is
\begin{eqnarray*}
\rho\left(t\right) & = & \left\vert \psi\left(t\right)\right\rangle \left\langle \psi\left(t\right)\right\vert .
\end{eqnarray*}
Then the reduced density matrix for charge qubit is calculated as
\begin{eqnarray}
\rho_{s}\left(t\right) & = & Tr\left(\left\vert \psi\left(t\right)\right\rangle \left\langle \psi\left(t\right)\right\vert \right))\nonumber \\
 & = & \alpha^{*}\alpha\left\vert 0\right\rangle \left\langle 0\right\vert +\langle\mu_{1}\left(t\right)\vert\mu_{0}\left(t\right)\rangle\alpha\beta^{*}\left\vert 0\right\rangle \left\langle 1\right\vert \nonumber \\
 &  & +\langle\mu_{0}\left(t\right)\vert\mu_{1}\left(t\right)\rangle\alpha^{*}\beta\left\vert 1\right\rangle \left\langle 0\right\vert +\beta^{*}\beta\left\vert 1\right\rangle \left\langle 1\right\vert . \label{eq:reduced density matrix}
\end{eqnarray}
As a measure of decoherence, the off-diagonal elements of the reduced
density matrix can be defined as decoherence factor~\cite{Sun93}
\[
D\left(t\right)=\left\vert \left\langle \mu_{1}\left(t\right)\vert\mu_{0}\left(t\right)\right\rangle \right\vert .
\]
To clarify the influence of the nonlinearity of nanomechanical resonator,
the decoherence factor will be calculated in the following.

Using the Baker-Hausdoff formulas, it is easily calculated that
\begin{eqnarray*}
U_{\chi}^{\dagger}(t)aU_{\chi}(t) & = & e^{i\chi ta^{\dagger}a}a,\\
U_{\chi}^{\dagger}(t)a^{\dagger}U_{\chi}(t) & = & a^{\dagger}e^{i\chi ta^{\dagger}a}.
\end{eqnarray*}
The decoherence factor becomes
\begin{align}
D\left(t\right) & =\vert\left\langle -\alpha_{1}e^{-i\Omega t}\right\vert U_{\chi}^{\dagger}(t)D^{\dagger}\left(\alpha_{1}\right)U_{\chi}(t)\label{d-factor}\\
 & U_{\chi}^{\dagger}(t)D\left(\alpha_{0}\right)U_{\chi}(t)\left\vert -\alpha_{0}e^{-i\Omega t}\right\rangle \vert\nonumber
\end{align}
where we have adopted $U_{\chi}(t)U_{\chi}^{\dagger}(t)=1$ for $U_{\chi}(t)=e^{-i\chi\left(a^{\dagger}a\right)^{2}t}$.

Taking the coherent state displacement operator
\[
D(\xi)=e^{-\frac{1}{2}|\xi|^{2}}\sum_{k,m=0}^{\infty}\frac{(\xi a^{\dagger})^{k}(-\xi^{*}a)^{m}}{k!m!}
\]
and
\[
\left(e^{i\chi ta^{\dagger}a}a\right)^{m}\left\vert \xi\right\rangle =\left(\xi\right)^{m}\left(e^{i\chi t}\right)^{\frac{m\left(m-1\right)}{2}}\left\vert \xi e^{im\chi t}\right\rangle
\]
into Eq.~(\ref{d-factor}), the decoherence factor is obtained
\begin{align}
D\left(t\right) & =\vert e^{-3\lambda^{2}}\sum_{k,m=0}^{\infty}\frac{\left(2\lambda^{2}\right)^{k+m}}{k!m!}\left(e^{i\Omega t}\right)^{k}\left(e^{-i\chi t}\right)^{\frac{k\left(k-1\right)}{2}}\nonumber \\
 & \left(e^{-i\Omega t}\right)^{m}\left(e^{i\chi t}\right)^{\frac{m\left(m-1\right)}{2}}e^{-\lambda^{2}-\lambda^{2}e^{-ik\chi t}e^{im\chi t}}\vert.\label{d-factor final}
\end{align}
Without the nonlinearity, $\chi=0$ and $\lambda=g/\omega$, the decoherence factor in Eq.~(\ref{d-factor final}) reduces into
\begin{equation}
D\left(t\right)  = e^{-8\frac{g^{2}}{\omega^{2}}\sin^{2}\left(\frac{1}{2}\omega t\right)}.\label{Gao05}
\end{equation}
Which is also obtained in the coupled system of charge qubit and microwave cavity.~\cite{Gao05}

To demonstrate the influence of the nonlinearity on the quantum decoherence, we consider it in the short-time limit. Here the condition $\chi t\ll1$ is assumed, we get $e^{i\chi t}\approx1+i\chi t$. Then the decoherence factor  in Eq.~(\ref{d-factor final}) is approximately
\begin{equation}
D\left(t\right)  = e^{-8\lambda^{2}\sin^{2}\left[\frac{1}{2}\left(\Omega+\lambda^{2}\chi+\frac{1}{2}\chi\right)t\right]}.\label{dfactor-simplify}
\end{equation}
Through some simple calculations, we have $\Omega+\lambda^{2}\chi+\frac{1}{2}\chi > \omega$ and $\lambda^{2} < g^{2}/\omega^{2}$. Seen in Eqs.~(\ref{Gao05}, \ref{dfactor-simplify}), the nonlinearity of nanomechanical resonator
increases the speed of oscillation and reduces the amplitude of oscillation in $D\left(t\right)$.

\section{Conclusions}

Considering the nonlinearity of nanomechanical resonator, the possibility
of implementing quantum information storage and quantum decoherence
of charge qubit are studied analytically. Using nonlinear Jaynes-Cummings
model, we find that the nonlinear nanomechanical resonator can not
be a quantum data bus. In addition, the equation~(\ref{probability})
shows that the nonlinearity does not destroy the revival of quantum
coherence of charge qubit. With the calculations of decoherence factor in Eqs.~(21-23), we demonstrate the influence of the nonlinearity of nanomechanical resonator on engineered decoherence of charge qubit.
It shows that the nonlinearity of nanomechanical resonator
affects the oscillation in decoherence factor $D\left(t\right)$.

\begin{acknowledgments}
We thank Dr. M. Hua and Dr. X. Xiao for helpful discussions.
\end{acknowledgments}

\end{document}